\documentstyle[12pt]{article}
\def\be{\begin{equation}}
\def\ee{\end{equation}}
\def\vk{{\bf k}}

\begin{document}

\begin{titlepage}

{}\hfill Preprint FIAN-TD/97-09

{}\hfill hep-ph/9706514

\begin{centering}
\vfill

{\bf ISOTOPIC SPIN CONSERVATION EFFECT\\
\vspace{0.6cm}
 IN BOSE-EINSTEIN CORRELATIONS}\\
\vspace{2cm}
I.V.ANDREEV\\
\vspace{0.3cm}
{\em P.N.Lebedev Physical Institute, Moscow 117924, Russia}\\
\vspace{1cm}

M.BIYAJIMA\\
\vspace{0.3cm}
{\em Department of Physics, Faculty of Liberal Arts,
Shinshu University,\\ Matsumoto 390, Japan}\\
\vspace{3cm}
{\bf Abstract}\\
\end{centering}
\vspace{0.3cm}

Bose-Einstein correlations of pions have been calculated in the presence
of both statistical and coherent pion production with isotopic spin
conservation taken into account.An additional contribution to pion
correlation function has been found arising due to isospin effect
\vspace{5cm}
\end{titlepage}

\vspace{3cm}
Bose-Einstein (BE) correlations of identical particles in multiple
production processes are extensively studied last years because they give an
information on the space-time region of interaction~\cite{{BGJ},{ABDS}}.
The basic effect is analogous to Hanbury Brown - Twiss (HBT)
interferometry in optics~\cite{HBT} and suggests statistical production
of the particles (mainly $\pi$ mesons). The possible presence of coherent
pionic component (for example, in the case of disoriented chiral condensate
formation) modifies the HBT effect. This modification in particle physics
was taken formerly in close analogy with optics~\cite{BMN}.

On the other hand, the pions (contrary to photons) are subject to
isotopic spin (and electric charge) conservation and so they can not
be emitted independently.
While the corresponding change of the statistical part is not essential
for large multiplicities (one must only ensure a symmetry between
isotopic spin components~\cite{APW93}), the coherent part changes substantially
when isotopic spin conservation is taken into account~\cite{{HS},{BSS},{A}}.
So we reconsider BE correlations of the pions in the presence
of both statistical and coherent components taking into account isotopic
spin conservation in the coherent part. That will result in appearance
of additional contribution to pionic correlation function.

We take annihilation (creation) operators $a_i$ of the pionic fields
as a sum of statistical ($b_i$) and coherent ($c_i$) parts,
\be
a_i({\vk})=b_i({\vk})+c_i({\vk}),\qquad   i=+,-,0.
\label{eq:1}
\ee
Normalizations can be chosen in such a way that invariant single-particle
inclusive cross-section for every kind of pions is
\be
\rho(\vk)=\frac{1}{\sigma}\frac{d\sigma}{dk}
=<a^{\dag}(\vk)a(\vk)>
=\rho^{ch}(\vk)+\rho^c(\vk)
\label{eq:2a}
\ee
with
\be
dk=d^3k/2E(2\pi)^3, \quad
\rho^{ch}(\vk)=<b^{\dag}(\vk)b(\vk)>,\quad  \rho^c(\vk)=<c^{\dag}(\vk)c(\vk)>
\label{eq:2b}
\ee
where the averaging brackets mean matrix element over pure pionic state
for $c$ operators and a statistical averaging for $b$ operators.
We use also the traditional chaoticity parameter
\be
p(\vk)=\rho^{ch}(\vk)/\rho(\vk)
\label{eq:3}
\ee
which gives share of the statistical (chaotic) component in the single-particle
 density.If statistical and coherent parts have different momentum
distributions then the chaoticity parameter depends on momentum $\vk$.

Consider now two-particle inclusive cross-section of identical pions
\begin{eqnarray}
\rho_2(\vk_1,\vk_2)=\frac{1}{\sigma}\frac{d^2\sigma}{dk_1dk_2}
=<a^{\dag}(\vk_1)a^{\dag}(\vk_2)a(\vk_1)a(\vk_2)>
\label{eq:4}
\end{eqnarray}
Splitting statistical average of the operator product into product
of pair expectations (with particle permutation) we get the correlation
function where BE statistics is taken into account:
\begin{eqnarray}
&C_2(\vk_1,\vk_2)=\rho_2(\vk_1,\vk_2)-\rho(\vk_1)\rho(\vk_2) \nonumber\\
&=|<b^{\dag}(\vk_1)b(\vk_2)>|^2
+2Re[<b^{\dag}(\vk_1)b(\vk_2)><c^{\dag}(\vk_2)c(\vk_1)> \nonumber\\
&+<c^{\dag}(\vk_1)c^{\dag}(\vk_2)c(\vk_1)c(\vk_2)>
-<c^{\dag}(\vk_1)c(\vk_1)><c^{\dag}(\vk_2)c(\vk_2)>
\label{eq:5}
\end{eqnarray}
for pions of every charge (here we omitted a small "surprising term"
in the correlation function~\cite{APW91}).

The first term in the right hand side of Eq.~\ref{eq:5} is the usual
effect of BE statistics (HBT effect), the second term represents
an interference effect of statistical and coherent production~\cite{BMN}
and the last two terms represent pure coherent contribution to the
correlation function.This last contribution vanishes in optics (for photons)
but survives in the case of coherent pion production due to pion
correlation arising through isotopic spin conservation.

Consider now the form of coherent contribution.
A typical ansatz for coherent pion state with fixed isospin is given
by zero isospin projection of the standard coherent state~\cite{HS,BSS}.
It has the form:
\be
|f>=\frac{1}{N}\int d\Omega\, exp\left\{-\frac{<n^c>}{2}
+\int dk f(\vk)\bf e\bf c^{\dag}(\vk)\right\}|0>
\label{eq:6}
\ee
where $N$ is normalization factor and
\be
<n^c>=\int dk\,|f(\vk)|^2
\label{eq:7}
\ee
Integration in Eq.~\ref{eq:6} is performed over directions of the unit
vector $\bf e$ in three-dimensional isotopic space,
\be
e_{\pm}=\frac{\sin\theta}{\sqrt2}e^{\pm i\phi},\qquad
e_0=\cos\theta
\label{eq:8}
\ee
for $\pi^{\pm}$ and $\pi^0$ mesons respectively and $|f(\vk)|^2$
gives momentum distribution of the coherently produced pions.
If we consider the space-time region in which the pions are produced
then $f(\bf k)$ is the mass-shell Fourier-transform of the space-time
region $f(x)$,
\be
f(\vk)=\int d^{4}x e^{-ikx}f(x)|_{k_0=E_k}
\label{eq:9}
\ee
This function is not necessarily real,
\be
f(\vk)=|f(\vk)|e^{i\varphi(\vk)}
\label{eq:10}
\ee
We suggest below that the average number of pions is large,
$<n^c>\gg1$, then
\be
N\approx8\pi^2/<n^c>
\label{eq:11}
\ee

Now it is necessary to calculate the matrix elements in the last two terms
in Eq.~\ref{eq:5} over pionic state of Eq.~\ref{eq:6}.
To do that we act by $c_i$-operators to the right and by $c^{\dag}_i$-operators
to the left.For large $<n^c>$ the calculation is greatly simplified
because in this case one may use steepest descent method in the coarse
of integration over spherical angles $\theta_1,\phi_1,\theta_2,\phi_2$
entering the states.Then we get the main contribution from the region
$\theta_1\approx\theta_2,\phi_1\approx\phi_2$ and reduce the matrix
elements to the average over single solid angle $\Omega$.The resulting
expressions for matrix elements take a simple form:
\be
<c^{\dag}_i(\vk)c_i(\vk)>=<n^c_i(\vk)>=\frac{1}{3}|f(\vk)|^2,\qquad
i=+,-,0
\label{eq:13}
\ee
\be
<c^{\dag}_i(\vk_1)c^{\dag}_i(\vk_2)c_i(\vk_1)c_i(\vk_2)>
=r_i|f(\vk_1)f(\vk_2)|^2
\label{eq:14}
\ee
with 
\be
r_+=r_-\approx\frac{2}{15},\qquad r_0\approx\frac{1}{5}
\label{eq:15}
\ee
It is traditionary in the literature to express BE correlations through
relative correlation functions
\be
C^{rel}_2(\vk_1,\vk_2)=\frac{\rho_2(\vk_1,\vk_2)-\rho(\vk_1)\rho(\vk_2)}
{\rho(\vk_1)\rho(\vk_2)}
\label{eq:16}
\ee
\be
d^{ch}(\vk_1,\vk_2)=\frac{<b^{\dag}(\vk_1)b(\vk_2)>}
{[\rho^{ch}(\vk_1)\rho^{ch}(\vk_2)]^{\frac{1}{2}}}
\label{eq:17}
\ee
Using above expressions we get from Eq.~\ref{eq:5}:
\begin{eqnarray}
C^{rel}_2(\vk_1,\vk_2)&=&p_1p_2|d^{ch}(\vk_1,\vk_2)|^2 \nonumber\\
&+&2\sqrt{p_1p_2(1-p_1)(1-p_2)}Re[d^{ch}(\vk_1,\vk_2)e^{i(\varphi_2-\varphi_1)}]
\nonumber\\&+&\beta(1-p_1)(1-p_2)
\label{eq:18}
\end{eqnarray}
with
$$
p_{1,2}=p(\vk_{1,2}),\quad \varphi_{1,2}=\varphi(\vk_{1,2})
$$
and
\be
\beta_{++}=\beta_{--}\approx\frac{1}{5},\quad \beta_{00}\approx\frac{4}{5}
\label{eq:19}
\ee
for $\pi^+\pi^+$,$\pi^-\pi^-$ and $\pi^0\pi^0$ mesons.In general,
coefficients $\beta_{ii}$ depend on total isospin of pionic system.

Therefore the presence of isotopic spin and its conservation
 leads to additional positive contribution 
to identical pion correlation dependent on the portion of the coherent
component.If the portion $1-p(\vk)$ does not decrease with increasing
of the momentum $\vk$ then this contribution leads to long range
correlations.Note that the coherent contribution to the
relative correlation function of Eq.~\ref{eq:18} is a multiplicative
function of particle momenta, not a function of momenta difference.
This feature may help to identify the coherent contribution in
experimental data on identical pion correlations.

\subsection*{Acknowledgments}
This work was supported in part by the JSPS Program on Japan-FSU
Scientists Collaboration. I.V.A. was also supported by Russian Fund
for Fundamental Research, grant 96-02-16210.

\end{document}